\documentclass[preprint,showpacs,lengthcheck,epsf]{revtex4}
\usepackage{xspace}
\usepackage{epsfig}
\usepackage{amsmath}
\usepackage{eufrak}
\begin{document}
%\psdraft
\title{Realization of wide electron slabs by polarization bulk doping in \\
graded III-V nitride semiconductor alloys}
\author{Debdeep Jena$^{1}$}
\author{Sten Heikman}
\author{Daniel Green}
\author{Ilan B. Yaacov}
\author{Robert Coffie}
\author{Huili Xing}
\author{Stacia Keller}
\author{Steve DenBaars}
\author{James S. Speck}
\author{Umesh K. Mishra}
\affiliation{Department of Electrical and Computer Engineering and Materials Department \\
            University of California, Santa Barbara \\
            CA, 93106}
\begin{abstract}
We present the concept and experimental realization of
polarization-induced bulk electron doping in III-V nitride
semiconductors.  By exploiting the large polarization charges in
the III-V nitrides, we are able to create wide slabs of high
density mobile electrons without introducing shallow donors.
Transport measurements reveal the superior properties of the
polarization doped electron distributions than comparable shallow
donor doped structures.  The technique is readily employed for
creating highly conductive layers in many device structures.
\end{abstract}
\footnotetext[1] {djena@engineering.ucsb.edu}
 \maketitle

%% Comment this out for single-column output:
%%
%% \twocolumn

%% This tells REVTeX to print out the numbers in the \pacs command:
%\draft

%%%%% Title:
%%
%%%%% Author:
%%
%%%%% Abstract:
%%
%%
%%%%%

%%%%% Physics and Astronomy Classsification Scheme (PACS)
%%        indexing codes:
%%

%%%%% Text of paper:
%%
%

Doping in semiconductors has been a much researched topic. The
traditional shallow `hydrogenic' doping technique is very well
understood and gainfully employed.  A good understanding of the
role of ionized dopant atoms on carrier scattering in
semiconductors led to the concept of modulation doping, which
improved low temperature carrier mobilities in quantum-confined
structures by many orders of magnitude \cite{pfeiffer}.

The last decade witnessed the emergence of the III-V nitrides as a
wide bandgap semiconductor with the property of large embedded
electronic polarization fields owing to the lack of inversion
symmetry in the crystal structure
\cite{bernardini},\cite{patrick}.  This property has been widely
exploited to make nominally undoped two-dimensional electron gases
(2DEGs) in AlGaN/GaN heterostructures, which had led to
high-electron mobility transistors (HEMTs) with record high
performance characteristics\cite{yifeng}.  The 2DEG at the
AlGaN/GaN interface of a III-V nitride heterostructure is formed
to screen the polarization dipole (with spontaneous and
piezoelectric contributions) in the thin epitaxial AlGaN cap
layer.  Surface donor-like states act as modulation dopants,
supplying electrons to form a dipole with the 2DEG at the
heterointerface\cite{ibbo}.

The discontinuity of polarization across an
Al$_{x}$Ga$_{1-x}$N/GaN heterojunction $\Delta
P_{hj}=P_{tot}^{AlGaN}(x)-P_{Sp}^{GaN}$ forms a fixed polarization
sheet charge at the heterojunction.  Grading the AlGaN/GaN
heterojunction over a distance should spread the positive
polarization sheet charge into a bulk three-dimensional
polarization background charge.  The charge profile is given by
the divergence of the polarization field, which changes only along
the (growth)z-direction ($N_{D}^{Pol}(z) = \nabla \cdot P =
\frac{\partial P(z) }{\partial z}$).  This fixed charge profile
will depend on the nature of the grading; a linear grade results
in an approximately uniform profile given by
$N_{D}^{Pol}(z)=\frac{P(z_{0})-P(0)}{z_{0}}$.  Here $P(z_{0})$ is
the polarization (spontaneous+piezoelectric) of
Al$_{x}$Ga$_{1-x}$N at the local Al composition at $z=z_{0}$.

This fixed background charge attracts free carriers from remote
donor-like states to satisfy Poisson's equation and charge
neutrality.  Figure 1 shows the schematic band diagrams and charge
profile showing the effect of linear grading of the
heterojunction.  The end result of the charge rearrangements makes
the polarization bulk charge act as a local donor with zero
activation energy.  The mobile three-dimensional electron slab
(3DES) thus formed should be usable just as bulk doped carriers.
However, removal of ionized impurity scattering should result in
higher mobilities.  Such polarization induced electron slabs
should in principle be similar to the modulation doped
three-dimensional electron slabs in modulation doped wide
parabolically graded quantum wells in the AlGaAs/GaAs system
\cite{shayegan}.  The mobile 3DES should not freeze out at low
temperature (as shallow donor doped bulk carriers do), and should
exhibit high mobilities at low temperatures.

To verify these concepts, five samples were grown by molecular
beam epitaxy (MBE).  Active nitrogen was provided by a rf-plasma
source.  High resistivity semi-insulating (SI) GaN on sapphire
grown by metal-organic chemical vapor deposition (MOCVD) was used
as templates.  For all five samples, a 100nm buffer MBE layer of
undoped (Ga-face) GaN was grown, followed by a different cap layer
for each. The cap layer for the five samples are described in
Table I. The top 100nm of sample 1 is bulk shallow donor doped
with Si (activation energy $E_{D}=20 meV$, and concentration $
N_{D}=10^{18}/cm^{3}$).  Samples 2,3 and 4 are linearly graded
AlGaN/GaN structures for studying polarization bulk doping; they
are graded from GaN to 10\%, 20\% and 30\% AlGaN respectively over
$z_{0}$=100nm.  Sample 5 is a 20nm Al$_{0.2}$Ga$_{0.8}$N/GaN which
houses a conventional 2DEG at the heterojunction.  Samples 1 and 5
are control samples.

Triple-crystal X-Ray diffraction data around the GaN (00.2) peak
of samples 1-4 is shown in Figure 2.  The data points match very
well with the theoretical solid curves\cite{xray} reflecting the
high degree of control of Al composition and growth rate in MBE.
Atomic force microscopy (AFM) of the sample surfaces revealed
step-flow growth and fully strained graded AlGaN surfaces.
Secondary Ion Mass Spectroscopy (SIMS) was performed on an extra
graded AlGaN layer sample specifically grown for that purpose. The
linearity of Al composition in the graded layer was revealed by
SIMS to be very accurately controlled.  It also revealed
background oxygen concentration in the MBE GaN layer to be
identical to the underlying MOCVD layer accompanied with a small
increase in the AlGaN layers.  Any background oxygen ( which acts
as a shallow donor in (Al)GaN ) may provide a small amount of
thermally activated carriers which can be frozen out at low
temperatures.

Temperature dependent (20-300K) Hall measurements were performed
on all the five samples.  Table I shows room temperature and 30K
Hall measurement data for all five samples.  The table includes
the free carrier density in the bulk GaN and polarization induced
3DES and 2DEG densities calculated by solving Schrodinger and
Poisson equations self consistently for samples 2-5.  The room
temperature sheet conductivity $\sigma=qn\mu$ is also shown.
Temperature dependent carrier densities and mobilities for samples
1,4,and 5 are plotted in Figure 3 for comparison.  Carriers in the
0-30\% graded AlGaN sample mimics the transport characteristics of
modulation doped 2DEGs and 3DESs characterized by a lack of
activation energy, leading to a temperature independent carrier
density.  Carriers in the bulk donor doped sample show the
characteristic freeze-out associated with the hydrogenic shallow
donor nature of Si in bulk GaN.  A fit to theoretical dopant
activation yielded an activation energy\cite{degenerate} $E_{D}=$
20 meV with a doping density (fixed by the Si flux in MBE)
$N_{D}=10^{18}/cm^{3}$.  The activation energy of Si closely
matches that reported by Gotz et. al \cite{gotz}.  2DEG carrier
mobilities (Sample 5) are higher than the shallow donor doped and
polarization doped carriers both at room temperature and low
temperatures.

The point of interest is the order of magnitude improvement of
carrier mobility at low temperatures for the polarization doped
3DESs over comparable donor doped samples.  In donor doped GaN,
thermally activated carriers freeze out with lowering of
temperature leading to less energetic electrons and less effective
screening.  This causes severe ionized impurity scattering,
lowering the mobility.  However, the removal of ionized impurity
scattering in the polarization doped structure, aided by the
complete lack of carrier freezeout at low temperatures results in
much improved mobilities.  It is not clear yet what limits the low
temperature mobility of polarization doped 3DESs.  Alloy disorder
scattering could be a strong candidate since the 3DES is housed in
a linearly graded disordered alloy potential.  There is also an
improvement of low temperature mobility with increasing carrier
density, which points towards possible Coulombic scattering from
surface donors, charged dislocations, and background shallow
impurities.  Dislocation scattering in the polarization doped 3DES
is also reduced at low temperatures as compared to donor doped
carriers owing to the degenerate nature of 3DES
carriers\cite{myfirstpaper}.  Scattering from disorder in
microscopic dipoles forming the graded alloy polarization charge
could also be a source of scattering\cite{mydipolepaper}. The
unanswered questions open up avenues for further work in transport
of polarization doped 3DESs.

Of special interest to device engineers is the room temperature
mobility, and especially the conductivity $\sigma = e n \mu$. From
Table I, we see that the room temperature charge-mobility product
of the polarization doped 3DES (Sample 3) is more than {\em
double} of that of the comparable donor doped sample (Sample 1).
Further, the trend with increasing alloy composition suggests that
the conductivity {\em increases} with increasing carrier density
(got by either grading to higher aluminum composition for the same
thickness, or decreasing the thickness for same grading
composition).  This trend has proved very useful for our design of
high conductivity layers required in many device structures,
especially in field effect transistors (FETs) and regrown ohmic
contacts.  The additional band-discontinuity achieved at a regrown
polarization-doped AlGaN contact serves as an efficient
hot-electron launcher from the source into the FET channel,
reducing the transit times.  The flexibility of polarization
doping by grading (by controlling alloy composition and/or graded
layer thickness independently) is an added attraction.  An
interesting extension would be the possibility of achieving
polarization doped p-type carriers with higher mobilities by
grading down from AlGaN for Ga-face III-V nitrides. This might
solve the problems associated with the high activation energy of
the commonly used acceptor (Mg) for GaN.  Our work presents the
first step towards realizing the proposed enhancement of base
conductivity in AlGaN/GaN heterojunction bipolar transistors by
exploiting the strong electronic polarization properties of the
III-V nitride semiconductors \cite{asbeck}.

In conclusion, we have demonstrated that polarization fields can
be engineered to achieve bulk doping as an attractive alternate
doping technique in III-V nitride semiconductors.  We demonstrate
improved conductivity of polarization doped layers over comparable
donor doped layers, and point out avenues where it may be
gainfully employed.

The authors would like to thank Patrick Waltereit, Arthur C.
Gossard and Herbert Kroemer for useful discussions.  Funding from
POLARIS/MURI (Contract monitor: C. Wood) is gratefully
acknowledged.

\pagebreak

%%%%% Bibliography
%%

%%
%%%%%
%%%%% Table

\pagebreak

\textbf{Table caption}

Table I : Sample structures and Hall measurement data for the five
samples.

\pagebreak

\textbf{Figure captions}

Figure 1 : Schematic band diagram and charge profile of
polarization bulk doping by spreading the heterojunction sheet
charge into a bulk charge.\\

Figure 2 : Triple crystal X-Ray diffraction data (dots) and
theoretical curve (solid line) for Samples 1-4 around the GaN
(00.2) peak.  The close agreement between theoretically predicted
and experimentally measured values indicates well controlled MBE growth.\\

Figure 3 : Temperature dependent carrier sheet densities and
mobility for a Polarization doped (Sample 4), Donor doped (Sample
1) and a 2DEG (Sample 5) structures.  Note the improvement in
mobility and the lack of carrier activation for the polarization
doped electrons compared to the donor doped electrons. \\

\begin{table}
\caption{}
\begin{center}
\begin{tabular}{|c|c|ccc|ccc|}
  \hline
  Sample & Cap layer & \multicolumn{3}{c|}{Hall sheet density($cm^{-2}$)} & \multicolumn{2}{c|}{Hall mobility ($\frac{cm^{2}}{V \cdot s}$)} & \multicolumn{1}{c|}{300K Conductivity} \\
  & & \multicolumn{1}{c|}{Theory} & \multicolumn{1}{c|}{30K} & \multicolumn{1}{c|}{300K} &  \multicolumn{1}{c}{30K} & \multicolumn{1}{c|}{300K} & \multicolumn{1}{c|}{($10^{-4} \Omega^{-1}$)} \\
  \hline
  \hline
  1 & 100nm Bulk Si doped GaN  &  \multicolumn{1}{c|}{-} & \multicolumn{1}{c|}{$7.3 \cdot 10^{11}$ } & \multicolumn{1}{c|}{$7.0 \cdot 10^{12}$} &  \multicolumn{1}{c|}{$139$} & \multicolumn{1}{c|}{$329$} & \multicolumn{1}{c|}{$2.3$} \\
  \hline
  2 & 100nm 0-10\% lin. gr. AlGaN &  \multicolumn{1}{c|}{$2.5 \cdot 10^{12}$} & \multicolumn{1}{c|}{$2.0 \cdot 10^{12}$} & \multicolumn{1}{c|}{$1.7 \cdot 10^{12}$} & \multicolumn{1}{c|}{$1441$} & \multicolumn{1}{c|}{$386$} & \multicolumn{1}{c|}{$0.7$} \\
  \hline
  3 & 100nm 0-20\% lin. gr. AlGaN &  \multicolumn{1}{c|}{$5.8 \cdot 10^{12}$} & \multicolumn{1}{c|}{$4.9 \cdot 10^{12}$} & \multicolumn{1}{c|}{$7.8 \cdot 10^{12}$}  & \multicolumn{1}{c|}{$2556$} & \multicolumn{1}{c|}{$598$} & \multicolumn{1}{c|}{$4.7$} \\
  \hline
  4 & 100nm 0-30\% lin. gr. AlGaN &  \multicolumn{1}{c|}{$9.0 \cdot 10^{12}$} & \multicolumn{1}{c|}{$9.1 \cdot 10^{12}$} & \multicolumn{1}{c|}{$8.9 \cdot 10^{12}$} &  \multicolumn{1}{c|}{$2605$} & \multicolumn{1}{c|}{$715$} & \multicolumn{1}{c|}{$6.4$} \\
  \hline
  5 & 20nm Al$_{0.20}$Ga$_{0.80}$N / GaN  &  \multicolumn{1}{c|}{$7.7 \cdot 10^{12}$} &  \multicolumn{1}{c|}{$7.7 \cdot 10^{12}$} & \multicolumn{1}{c|}{$7.8 \cdot 10^{12}$} & \multicolumn{1}{c|}{$5644$} & \multicolumn{1}{c|}{$1206$} & \multicolumn{1}{c|}{$9.4$} \\
  \hline
  \hline
\end{tabular}
\end{center}
\label{InGaNPLstructTab}
\end{table}

%%%%% Figures
%%

% -----------------------------------------------------------------
\begin{figure}[h!]
\centerline{\epsfig{file=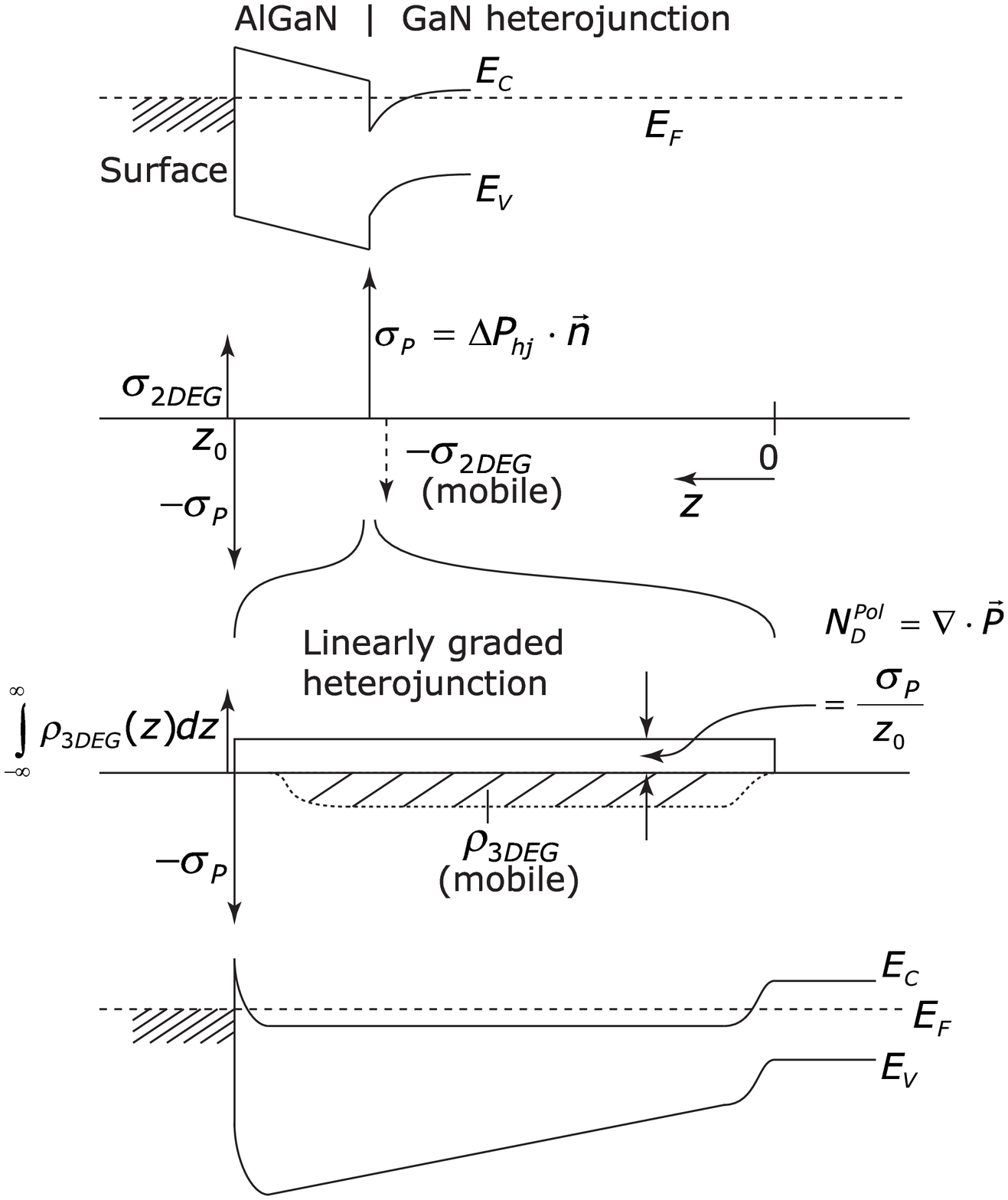,width=16cm}}
\caption{(D. Jena \textit{et. al.})} \label{banddiagrams}
\end{figure}

% -----------------------------------------------------------------
\begin{figure}[h!]
\centerline{\epsfig{file=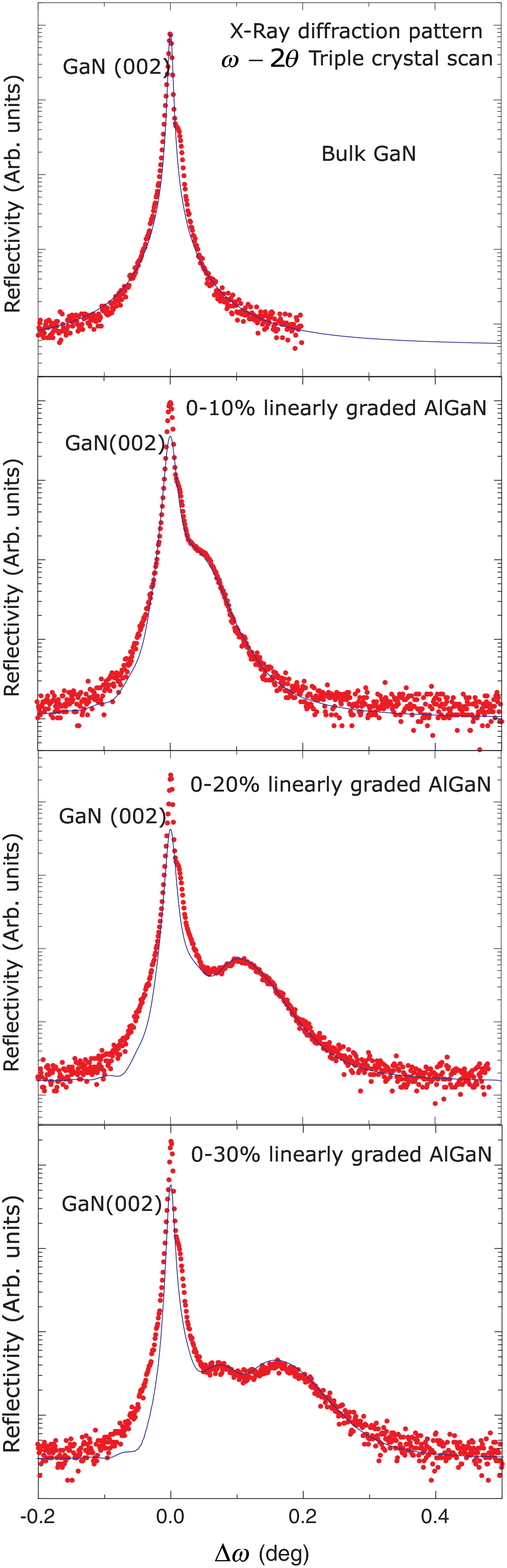,width=6cm}} \caption{(D. Jena
\textit{et. al.})} \label{XRay}
\end{figure}

% -----------------------------------------------------------------
\begin{figure}[h!]
\centerline{\epsfig{file=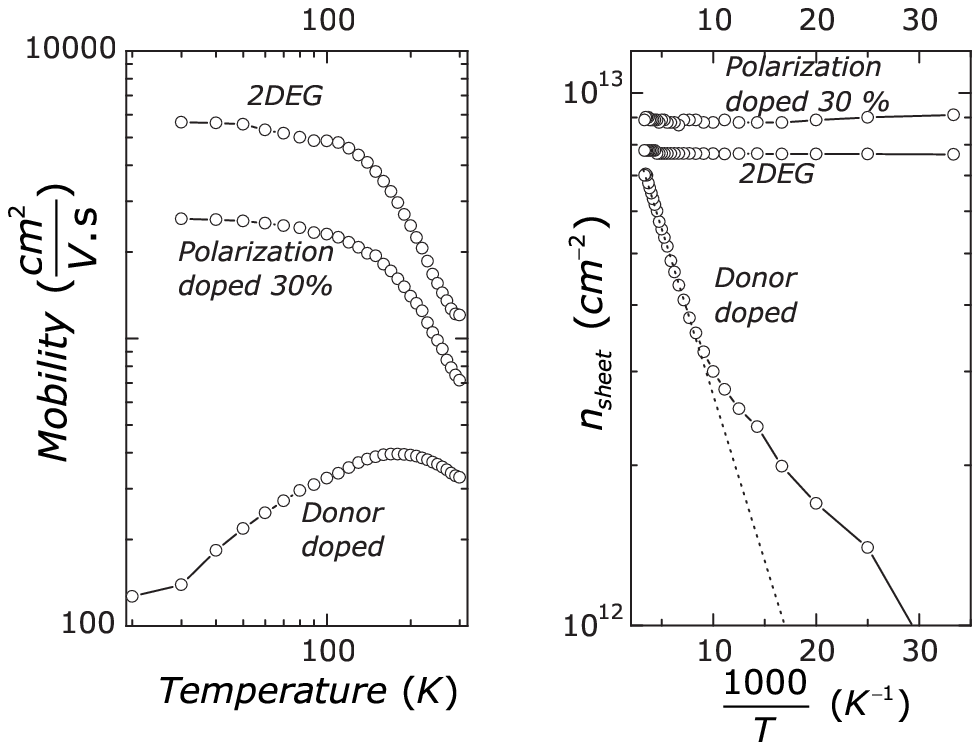,width=16cm}}
\caption{(D. Jena \textit{et. al.})} \label{TDepHall}
\end{figure}

%% FigurePageEnd

\end{document}